\documentclass[10pt,conference]{IEEEtran}
\usepackage{cite}
\usepackage{amsmath,amssymb,amsfonts}
\usepackage{algorithmic}
\usepackage{graphicx}
\usepackage{textcomp}
\usepackage{xcolor}
\usepackage[hyphens]{url}

\usepackage{fancyhdr}
\usepackage[hidelinks]{hyperref}
\usepackage{braket}
\usepackage{subcaption}
\usepackage{booktabs}
\usepackage{makecell}
\usepackage{float}
\usepackage{enumitem}
\makeatletter
\newcommand{\linebreakand}{
  \end{@IEEEauthorhalign}
  \hfill\mbox{}\par
  \mbox{}\hfill\begin{@IEEEauthorhalign}
}
\makeatother

\usepackage{tikz}
\usetikzlibrary{svg.path}
\usepackage{scalerel}

\begin{document}
\definecolor{orcidlogocol}{HTML}{A6CE39}
\tikzset{
  orcidlogo/.pic={
    \fill[orcidlogocol]
    svg{M256,128c0,70.7-57.3,128-128,128C57.3,256,0,198.7,0,128C0,57.3,57.3,0,128,0C198.7,0,256,57.3,256,128z};
    \fill[white] svg{M86.3,186.2H70.9V79.1h15.4v48.4V186.2z}
    svg{M108.9,79.1h41.6c39.6,0,57,28.3,57,53.6c0,27.5-21.5,53.6-56.8,53.6h-41.8V79.1z
    M124.3,172.4h24.5c34.9,0,42.9-26.5,42.9-39.7c0-21.5-13.7-39.7-43.7-39.7h-23.7V172.4z}
    svg{M88.7,56.8c0,5.5-4.5,10.1-10.1,10.1c-5.6,0-10.1-4.6-10.1-10.1c0-5.6,4.5-10.1,10.1-10.1C84.2,46.7,88.7,51.3,88.7,56.8z};
  }
}

\newcommand\orcidicon[1]{\textsuperscript{\href{https://orcid.org/#1}{\mbox{\scalerel*{
          \begin{tikzpicture}[yscale=-1,transform shape]
            \pic{orcidlogo};
          \end{tikzpicture}
}{|}}}}}

\title{Stalls and Spequlation: Pipelined Execution for Fault Tolerant Quantum Computation}
\author{
\IEEEauthorblockN{
Aditi Awasthi\IEEEauthorrefmark{3}\orcidicon{0009-0006-7768-8503},
Gokul Subramanian Ravi\IEEEauthorrefmark{4}\orcidicon{0000-0002-2334-2682},
Jonathan Mark Baker\IEEEauthorrefmark{3}\orcidicon{0000-0002-0775-8274}
}
\IEEEauthorblockA{
\IEEEauthorrefmark{3}Electrical and Computer Engineering, The University of Texas at Austin\\
\IEEEauthorrefmark{4}Computer Science and Engineering, University of Michigan\\
\IEEEauthorrefmark{1}\href{mailto:aditiawasthi@utexas.edu}{aditiawasthi@utexas.edu}
}
}
\maketitle

\begin{abstract}
Fault-tolerant quantum computation requires the coordinated action of three distinct systems: classical control logic, quantum hardware, and classical error decoders. Current scheduling models treat logical operations as atomic, hiding the fact that these subsystems operate sequentially and spend significant time idle. We present a pipelined execution framework that decomposes each logical operation into its component stages i.e. Control, Execute, and Decode. Building on this, we discuss some speculation strategies that allow successor operations to begin processing before their predecessors have completed decoding. We evaluate our framework on several common benchmarks and show that pipelining with speculation reduces total pipeline steps by 20-40\% compared to a no-speculation baseline. The most aggressive strategy consistently outperforms conservative alternatives, even though partial rollback is needed at times, because the per-rollback penalty is small relative to the parallelism gained. We further show that speculation facilitates load balancing by distributing work more evenly across the heterogeneous subsystems of a fault-tolerant quantum computer, converting idle time into useful computation while also saving on execution time.
\end{abstract}


\section{Introduction}
Quantum error correction is essential for realizing the computational potential of quantum computers~\cite{terhal2015quantum, fowler_2012}. The surface code has emerged as a leading candidate due to its high fault-tolerance threshold and its compatibility with two-dimensional nearest-neighbor hardware layouts. Lattice surgery, introduced by Horsman et al.~\cite{horsman2012surface}, provides a practical framework for performing logical operations on surface code patches through merge and split operations, and has since become the dominant paradigm for surface-code-based fault-tolerant quantum computation~\cite{Litinski_2019_lattice_surgery, fowler2019lowoverhead}.
 
Within this paradigm, the non-Clifford $T$ gate presents a well-known bottleneck. Because the surface code does not natively support transversal $T$ gates, they must be implemented via gate teleportation using auxiliary resource states known as magic states~\cite{Bravyi_2005}. Producing these states requires dedicated distillation circuits that consume a significant fraction of a fault-tolerant system's physical qubits~\cite{Litinski_2019_distillation}. The rate at which factories can supply magic states directly constrains how fast a quantum program can execute. Separately, the decoder backlog problem (in which classical decoding cannot keep pace with syndrome generation) has attracted significant attention, with solutions based on parallel window decoding~\cite{Skoric_2023}, and more recently, speculative prediction of inter-window dependencies~\cite{viszlai2025swiper}.

More broadly, the existing literature treats logical operations as atomic units with fixed time costs measured in code cycles---for instance, Beverland et al.~\cite{qre2} and Litinski~\cite{Litinski_2019_lattice_surgery} model each logical operation as a fixed-cost block, leaving no room to overlap the classical and quantum phases of adjacent operations. This abstraction is useful for asymptotic resource estimation, but it obscures a fundamental architectural reality: executing a single logical operation is not a monolithic event. It involves the coordinated action of multiple, heterogeneous computational subsystems including classical controllers, quantum hardware, and classical decoders, each with distinct latency characteristics that vary across hardware platforms, algorithm choices, and system configurations. When these subsystems operate in strict lockstep, each one is idle while the others work. The resulting underutilization is a significant and avoidable source of inefficiency that current scheduling frameworks do not directly address at the logical operation scheduling level.

Classical computer architecture solved a structurally analogous problem decades ago with instruction pipelining and speculative execution. We show that these techniques have direct and productive analogues in the quantum setting, with the additional advantage that the algebraic structure of the Clifford group makes certain speculation provably safe.
 
In this paper, we present a pipelined execution framework for fault-tolerant quantum programs that decomposes each logical operation into its constituent stages and overlaps them across independent operations in the program's dependency graph. We introduce three speculation strategies that allow successor operations to begin processing before their predecessors have fully completed, distributing computational work more evenly across the system's heterogeneous subsystems.

Our contributions are:
\begin{enumerate}
    \item A preliminary pipelined execution model that decomposes logical operations into \textit{Control} $\rightarrow$ \textit{Execute} $\rightarrow$ \textit{Decode} stages and exposes the utilization imbalance created by sequential scheduling across heterogeneous subsystems.
    \item Three speculation strategies: Aggressive, Commute-Aware (provably rollback-free for Z-type successors), and T-Cautious, with arguments for their correctness.
    \item An evaluation on benchmark circuits showing 20-40\% reduction in number of pipeline steps\footnote{A \emph{pipeline step} is a single stage (Control, Execute, or Decode) of one logical operation. In conventional models where operations are atomic, one logical operation $\equiv$ one code cycle; in our model it occupies 3 pipeline steps executed sequentially. Without overlap (i.e. sequential execution without speculation), total pipeline steps is simply $3\times$ the conventional cycle count.} and quantifying the utilization smoothing effect across subsystems.
\end{enumerate}

\section{Background}
\subsection{The Scheduling Problem}
\label{sec:scheduling}
Current resource estimation frameworks model each logical operation as an atomic unit with a fixed cost in code cycles~\cite{Litinski_2019_lattice_surgery, qre2}. In each code cycle, exactly one subsystem is active: the classical controller, the quantum hardware, or the decoder, while the other two sit idle. This idle time is not an inherent cost of fault tolerance; it arises from the decision to treat each operation as an indivisible block. The decoder backlog problem~\cite{terhal2015quantum} is one manifestation, but the higher-level scheduling problem: how to overlap the classical, quantum, and decoding phases of \emph{different} logical operations has received comparatively little attention. Our work addresses this gap.
 
\subsection{A Classical Precedent}
Classical processor design solved a structurally identical problem: multiple specialized subsystems idling in alternation, through instruction pipelining and speculative execution i.e. beginning work on subsequent instructions before prior ones complete. Modern processors achieve branch prediction accuracies exceeding 95\%, making speculation overwhelmingly profitable despite occasional rollback.
Table~\ref{tab:analogy} summarizes the correspondence. The quantum setting is in one respect \emph{more} favorable: the Clifford group's algebraic structure guarantees that a large class of successor operations commute with injection corrections, making speculation over them provably safe with no rollback ever needed (Section~\ref{sec:injection}).

\subsection{Magic State Injection}
\label{sec:injection}
The Clifford gate set $\{H, S, \text{CNOT}\}$ is not universal; by the Gottesman-Knill theorem, Clifford circuits can be efficiently simulated classically. Adding the $T$ gate yields the Clifford+$T$ gate set, which \emph{is} universal~\cite{Bravyi_2005}. This makes the $T$ gate the primary source of scheduling overhead in fault-tolerant systems.

\begin{figure}[!h]
     \centering
     \includegraphics[width=0.7\linewidth]{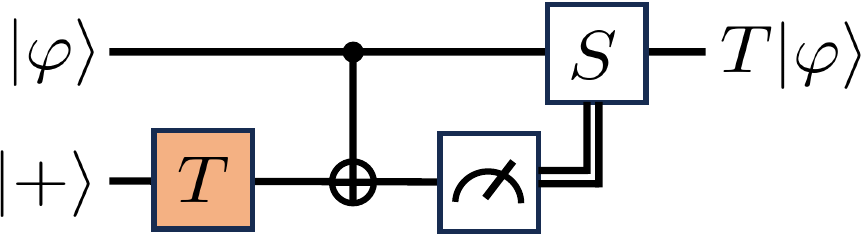}
     \caption{Magic state injection via gate teleportation. A magic state $\ket{M} = T\ket{+}$ is consumed through a CNOT interaction, followed by measurement. With probability $\frac{1}{2}$, an $S$ correction must be applied to the control qubit before subsequent operations can proceed.}
     \label{fig:t_injection}
\end{figure}

Clifford gates can be implemented natively via lattice surgery~\cite{horsman2012surface, Litinski_2019_lattice_surgery}. The $T$ gate must instead be realized through gate teleportation: a magic state $\ket{M} = T\ket{+}$ is consumed through a CNOT, followed by a measurement whose outcome determines whether an $S$ or $S^\dagger$ correction is required~\cite{Bravyi_2005, Litinski_2019_distillation} (see Figure~\ref{fig:t_injection}).

\begin{table}[!t]
\centering
\caption{Structural analogy between classical pipelined processors and pipelined fault-tolerant quantum execution}
\label{tab:analogy}
\begin{tabular}{p{0.4\linewidth} p{0.5\linewidth}}
\hline
\textbf{Classical Processor} & \textbf{Fault-Tolerant QC} \\
\hline
Instruction pipeline & Logical operation pipeline \\
Fetch $\to$ Decode $\to$ Execute $\to$ Writeback & classical pre-processing $\to$ quantum execution $\to$ classical decoding \\
Functional units (ALU, FPU) & Magic state factories \\
Conditional branch & $T$ gate measurement outcome \\
Speculative execution & Executing successors before decoding completes (this work) \\
Branch prediction (probabilistic, ${\sim}95\%$ accurate) & Commutativity analysis (deterministic for Clifford successors) \\
Branch misprediction penalty & Fixup cost on bad injection if speculation was too aggressive \\
Pipeline flush on mispredict & Undo $\to$ Fixup $\to$ Redo recovery \\
Data hazard stall & Waiting for magic state or correction resolution \\
\hline
\end{tabular}
\end{table}

This process is inherently non-deterministic. The measurement outcome is uniformly random, which means an $S$ or $S^\dagger$ correction must be applied to the control qubit with probability 50\%. The correction is itself a Clifford gate, but it creates a data dependency. Subsequent operations on the affected qubit must wait until the measurement outcome is known and any required correction has been applied, creating a stall in the computation (in sequential execution model).

The correction structure has an important algebraic property that is central to our speculation framework. Because the correction is a Clifford gate drawn from $\{S, S^\dagger\}$, any gate implementing a rotation about the $Z$-axis (including $Z$, $S$, $S^\dagger$, $T$, and $T^\dagger$) commutes with it, meaning computation can proceed without waiting for the correction outcome. Gates outside this set, such as Hadamard, do not commute with the correction in general, and executing them before the correction is resolved may require rollback. We additionally note that $Z$-type corrections on the \emph{control} qubit of a CNOT \emph{do} commute through, and we exploit this behavior.

\section{System Design}
Traditional conservative scheduling approaches require operations to wait for complete predecessor completion before beginning execution, creating artificial serialization that underutilizes available system resources and extends overall execution time. We apply two classical techniques to this setting: \emph{pipelining}, which decomposes each logical operation into overlapping stages, and \emph{speculation}, which allows successors to begin before predecessors have fully resolved. Our goal is to demonstrate that the pipelining and speculation paradigm is viable and beneficial in this setting, motivating further exploration.

\begin{figure*}[!t]
    \centering
\includegraphics[width=0.85\linewidth,height=4cm]{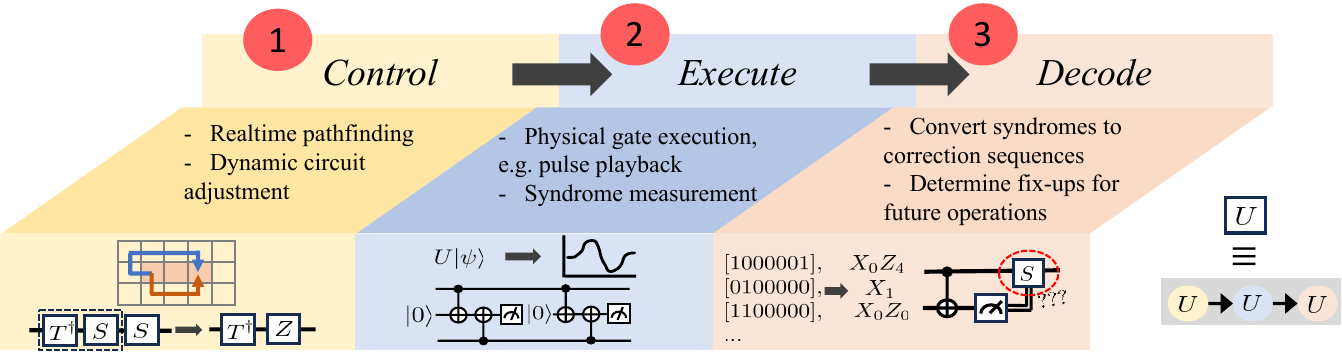}
    \caption{Three-stage execution pipeline for a logical operation. The \textit{Control} stage handles classical pre-processing, the \textit{Execute} stage performs physical gate application and syndrome measurement, and the \textit{Decode} stage runs classical error decoding}
    \label{fig:detailed_pipeline}
\end{figure*}

\subsection{Pipelining}
We model quantum gate execution as three pipeline stages: \textbf{control preparation}, \textbf{gate execution}, and \textbf{error syndrome decoding} (Figure~\ref{fig:detailed_pipeline}). These stages can have different durations depending on the hardware technology and the specific operation. In our model, $T$, $T^\dagger$, and CNOT gates occupy two steps per pipeline stage due to their additional coordination overhead, while all other gates occupy one step per stage. 
We weight Paulis (X, Z) identically to H and S to enable a static schedule that isolates the effect of pipelining. Although Pauli frame tracking could in principle make X and Z cheaper, the stalls our speculation strategies target arise from T-gate decode latency rather than from Pauli execution, so the comparison between strategies is insensitive to this modeling choice.
This abstraction captures the temporal structure of gate execution and the separation between classical and quantum actions associated with each gate.

\subsubsection{Control Stage}
This stage encompasses classical pre-processing tasks that can vary substantially in complexity. For simple Pauli (X, Z) or Hadamard operations, it may simply involve selecting appropriate control pulses and qubit addresses~\cite{dalvi2024one}. 
However, for more complex operations, it may include commuting gates to optimize circuit depth, adaptive circuit readjustment based on previous measurement outcomes~\cite{wang2024optimizing, Litinski_2019_lattice_surgery}, and executing real-time path-finding algorithms in systems with dynamic connectivity constraints~\cite{Sethi_2025}.

CNOT gates require coordination between multiple qubits and may involve routing through intermediaries in connectivity-constrained architectures. T gates execution requires coordination with magic state factories, resource allocation decisions, and potential queuing management when magic states are unavailable. This may lead to stalling and insertion of corrective operations at runtime due to injection failures. Thus, the circuit structure evolves dynamically, necessitating realtime rescheduling of downstream operations.

\subsubsection{Execute Stage}
This stage represents the application of physical gate operations on the target qubit(s), followed by syndrome extraction (the latter being a sequence of physical CNOT gates and measurements). Its duration is hardware-dependent: single-qubit gates complete in nanoseconds on superconducting platforms and microseconds on trapped-ion systems~\cite{Krantz_2019, Bruzewicz_2019, Slussarenko_2019}, while CNOT and T gates require additional time due to multi-qubit coordination and magic state consumption respectively.

\subsubsection{Decode Stage}
This stage is purely classical. It processes the syndrome measurements collected during execution, runs a decoding algorithm (e.g., minimum-weight perfect matching~\cite{wu_2023fusionblossomfastmwpm} or union-find~\cite{liyanage_2024}), and determines corrections needed for subsequent operations. Decoding latency can range from sub-microsecond for FPGA lookup-table decoders to tens of microseconds for more sophisticated algorithms, and grows superlinearly with code distance~\cite{ravi_2022}. The bandwidth available to transmit data from the surface code to the classical processor is also a critical factor, as syndrome data from all code patches must be processed in parallel.
When the decoder cannot keep pace with syndrome generation, a \emph{backlog} accumulates~\cite{terhal2015quantum,Skoric_2023}, precisely the kind of bottleneck that pipelining helps alleviate by overlapping decode work for one operation with control and execute work for others.\\

In this paper, we model all three pipeline stages as having equal duration (a 1:1:1 ratio), representing a balanced system design target. This simplification isolates the effect of scheduling strategy from hardware-specific timing asymmetries. In practice, some control stage computation should be started as early as possible (e.g., pre-computing routes), but other components like gate updates might rely on just-in-time information from recent measurements. Under this model, a single logical operation occupies three pipeline steps (one per stage), so the \textit{no-speculation} baseline step count is approximately $3\times$ the circuit's critical-path depth in logical operations.

\subsection{Speculative Execution}
\label{sec:spec}
\begin{figure}[t]
\centering
    \includegraphics[width=0.9\linewidth,height=7.1cm]{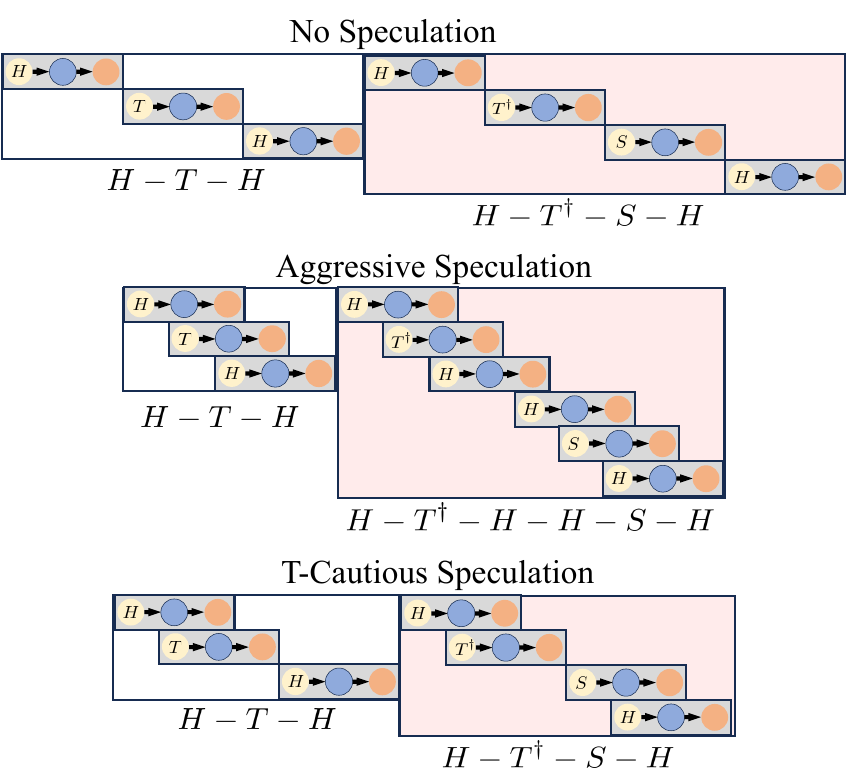}
    \caption{Gate progression through \textit{Control} (yellow), \textit{Execute} (blue), and \textit{Decode} (orange) stages. Left column: no injection error occurs. Right column: injection failure occurs, requiring correction. Each row shows a different scheduling strategy.}
    \label{fig:specstrat}
\end{figure}

Speculation allows successor operations to enter the pipeline before their predecessors have fully completed, reducing execution time and smoothing resource demand across cycles (an effect we quantify in Section~\ref{sec:utilization}). The fundamental challenge in aggressive pipelining and speculation stems from the need for post-measurement corrections if the gate application is erroneous. After gates complete execution, the decoding process may reveal that a fix-up is required. However, the application of these corrective gates can alter the computational context for subsequent operations that may have already begun speculative execution based on the assumption that no corrections would be needed (e.g. Figure~\ref{fig:specstrat}).

\textbf{Misprediction penalty.} When a T gate's decode indicates a bad injection, any speculatively advanced successor that does not commute with the $S/S^\dagger$ correction must be rolled back via an \emph{Undo $\rightarrow$ Fixup $\rightarrow$ Redo} sequence. This penalty is avoided entirely when the successor commutes with the correction (Section~\ref{sec:injection}), and is the key cost that distinguishes our three strategies.

We evaluate three speculation strategies that differ in the conditions under which a successor operation is permitted to enter the pipeline.

\subsubsection{\textbf{Aggressive Strategy}}
A successor operation may enter the \textit{Control} stage as soon as all its predecessors have entered their \textit{Execute} stage, regardless of gate type. If the predecessor's decode reveals a bad injection and the successor does not commute with the correction, the Undo $\rightarrow$ Fixup $\rightarrow$ Redo recovery protocol is invoked. This is the most aggressive strategy and achieves the highest parallelism at the cost of occasional rollback overhead.
\subsubsection{\textbf{Commute-Aware Strategy}}
A successor may enter the \textit{Control} stage early (before its predecessor $T$ or $T^\dagger$ is done with its \textit{Decode} stage) only if it commutes with the possible $S/S^\dagger$ correction from its predecessor. Concretely, speculation is permitted when the successor is a Z-type gate, or when the predecessor's qubit is the \textit{control} qubit of a subsequent CNOT. For non-commuting successors, the strategy falls back to waiting for the predecessor's decode to complete. This eliminates rollback for the commuting cases while still permitting speculation where it is provably safe.
\subsubsection{\textbf{T-Cautious Strategy}}
Speculation is permitted for all successors except when a predecessor $T$ or $T^\dagger$ gate has not yet completed decoding. This recognizes that $T$ gates are a source of non-deterministic corrections and restricts speculation accordingly, maintaining parallelism for Clifford-only dependency chains while avoiding all T gate-related misprediction penalties.\\

\noindent\textbf{Correctness}: All three strategies preserve the logical unitary implemented by the circuit. For the \textit{Aggressive strategy}, correctness follows from the recovery protocol: any speculatively advanced operation that conflicts with a correction is undone and re-executed with the correction in place, restoring the operation ordering to one consistent with the original DAG. For \textit{Commute-Aware}, the commutation property guarantees that Z-type successors produce the same result regardless of whether the correction is applied before or after them; the correction is inserted into the DAG after the commuting successor and applied when reached. For \textit{T-Cautious}, T predecessors are never speculated past until decoded, so no conflict can arise.

\begin{table}[!t]
\centering
\caption{Benchmark circuit characteristics after decomposition into Clifford+$T$ via GridSynth}
\label{tab:benchmarks}
\small
\begin{tabular}{lrrrrr}
\toprule
Benchmark & Gates & $T$/$T^\dagger$ & $T$ frac. & Depth \\
\midrule
\texttt{adder\_n10} & 138 & 48 & 35\% & 99 \\
\texttt{adder\_n4} & 23 & 8 & 35\% & 11 \\
\texttt{basis\_test\_n4} & 8712 & 3421 & 39\% & 3850 \\
\texttt{bell\_n4} & 6227 & 2470 & 40\% & 2025 \\
\texttt{bigadder\_n18} & 274 & 96 & 35\% & 152 \\
\texttt{dnn\_n2} & 4017 & 1581 & 39\% & 2418 \\
\texttt{fredkin\_n3} & 19 & 7 & 37\% & 11 \\
\texttt{ipea\_n2} & 4419 & 1745 & 39\% & 3022 \\
\texttt{ising\_n10} & 35577 & 14027 & 39\% & 6030 \\
\texttt{ising\_n26} & 12734 & 5014 & 39\% & 847 \\
\texttt{knn\_n25} & 12351 & 4853 & 39\% & 906 \\
\texttt{linearsolver\_n3} & 2431 & 958 & 39\% & 1426 \\
\texttt{multiplier\_n15} & 476 & 204 & 43\% & 234 \\
\texttt{multiply\_n13} & 95 & 36 & 38\% & 39 \\
\texttt{pea\_n5} & 6476 & 2554 & 39\% & 4258 \\
\texttt{qaoa\_n3} & 1201 & 474 & 39\% & 787 \\
\texttt{qft\_n4} & 1651 & 653 & 40\% & 1219 \\
\texttt{qpe\_n9} & 4912 & 1946 & 40\% & 3047 \\
\texttt{qram\_n20} & 3506 & 1384 & 39\% & 1118 \\
\texttt{quantumwalks\_n2} & 3995 & 1578 & 39\% & 2428 \\
\texttt{sat\_n11} & 3450 & 1374 & 40\% & 2226 \\
\texttt{seca\_n11} & 205 & 54 & 26\% & 68 \\
\texttt{shor\_n5} & 3246 & 1275 & 39\% & 2601 \\
\texttt{simon\_n6} & 42 & 14 & 33\% & 27 \\
\texttt{square\_root\_n18} & 1971 & 780 & 40\% & 1292 \\
\texttt{swap\_test\_n25} & 12372 & 4827 & 39\% & 904 \\
\texttt{toffoli\_n3} & 18 & 7 & 39\% & 12 \\
\texttt{variational\_n4} & 8121 & 3168 & 39\% & 4270 \\
\texttt{vqe\_n4} & 7311 & 2863 & 39\% & 2238 \\
\texttt{wstate\_n27} & 10229 & 3992 & 39\% & 5319 \\
\texttt{wstate\_n3} & 2608 & 1032 & 40\% & 1801 \\
\bottomrule
\end{tabular}
\end{table}

\section{Results and Observations}
\label{sec:results}
We evaluate on circuits from QASMBench~\cite{li2022qasmbenchlowlevelqasmbenchmark}, decomposed into the Clifford+$T$ gate set using GridSynth~\cite{ross_2016_gridsynth}. Table~\ref{tab:benchmarks} summarizes the benchmark characteristics. Each speculation strategy is compared against a pipelined no-speculation baseline, with unbounded magic state availability.

\begin{figure*}[!t]
    \centering
    \includegraphics[width=1\linewidth,height=4.7cm]{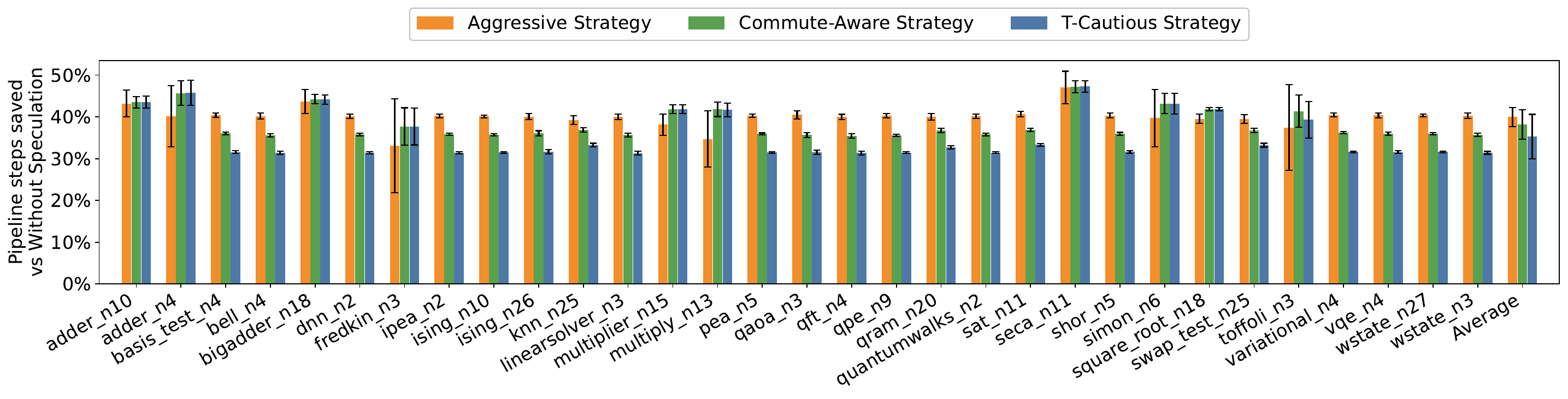}
    \caption{Percentage savings in pipeline steps of each speculation strategy relative to the no-speculation baseline. Aggressive speculation achieves the largest savings on most benchmarks}
    \label{fig:diffspecstrategies}
\end{figure*}

\subsection{Cycle Reduction} \label{sec:cycles}
Figure~\ref{fig:diffspecstrategies} shows the percentage cycle savings of each strategy relative to the no-speculation baseline.

\textbf{Speculation consistently reduces cycle time.} Every strategy achieves positive savings on all benchmarks; there is no circuit where speculation hurts, confirming that the pipelining benefit is robust.

\textbf{Aggressive speculation achieves the largest savings on almost all benchmarks.} On circuits with deep T-gate dependency chains (\texttt{ising\_n26}, \texttt{knn\_n25}), it achieves ${\sim}40\%$ reduction in pipeline steps. One might expect the aggressive strategy to suffer due to the 50\% injection failure rate. We measured $p_{\text{non-commute}}$ i.e. the fraction of T gate successors that do \emph{not} commute with the $S/S^\dagger$ correction across all benchmarks and found it ranges from 49\% to 100\% (mean: 61\%). This yields an effective rollback rate of $\frac{1}{2} \times p_{\text{non-commute}} \approx 24$-$50\%$, which is substantial. Yet, the aggressive strategy dominates. The reason is that the \emph{cost} of an Undo $\rightarrow$ Fixup $\rightarrow$ Redo sequence is small relative to the \emph{benefit} of speculating: each successfully speculated operation saves multiple idle pipeline steps that would otherwise be spent waiting for the predecessor's Decode stage to complete. The savings accumulate globally, while each individual rollback penalty is local. The net effect is decisively positive. However, the aggressive strategy exhibits higher variance across trials than the conservative alternatives, as visible in the large error bars of Figure~\ref{fig:diffspecstrategies}. This is expected since rollback is stochastic, and different trials will produce different correction patterns. In practice, applications requiring predictable latency may prefer \textit{Commute-Aware} or \textit{T-Cautious} for their tighter variance bounds.

\textbf{Commute-Aware and T-Cautious perform similarly.} These two strategies produce comparable pipeline step counts across most benchmarks, with Commute-Aware holding a slight edge. Given the high $p_{\text{non-commute}}$ values, both strategies are conservative in similar ways: they refuse to speculate past T gates in most cases. Commute-Aware gains a modest advantage by permitting speculation in the ${\sim}39\%$ of successor edges (on average) that do commute, but this advantage is limited because these tend to be Z-type gates that are already fast to execute.

\textbf{Benefits scale with circuit size but remain proportionally significant at small scale.} Absolute savings are naturally larger for bigger circuits, but even small benchmarks exhibit meaningful proportional reduction. This indicates that the benefit is structural: it arises from the pipeline overlap itself.

\subsection{Utilization Smoothing} \label{sec:utilization}
\begin{figure*}
  \centering
  \begin{subfigure}{\textwidth}
    \centering
        \includegraphics[width=0.35\linewidth]{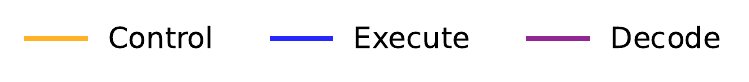}
    \end{subfigure}\\
  \begin{subfigure}{\textwidth}
    \includegraphics[height=3.25cm,width=\columnwidth]{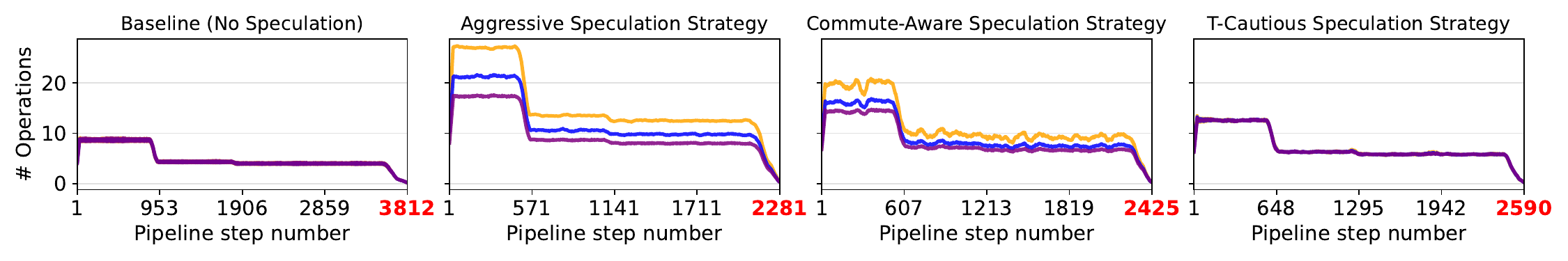}
    \caption{\texttt{ising\_n26}}
  \end{subfigure}
  \begin{subfigure}{\textwidth}
    \includegraphics[height=3cm,width=\columnwidth]{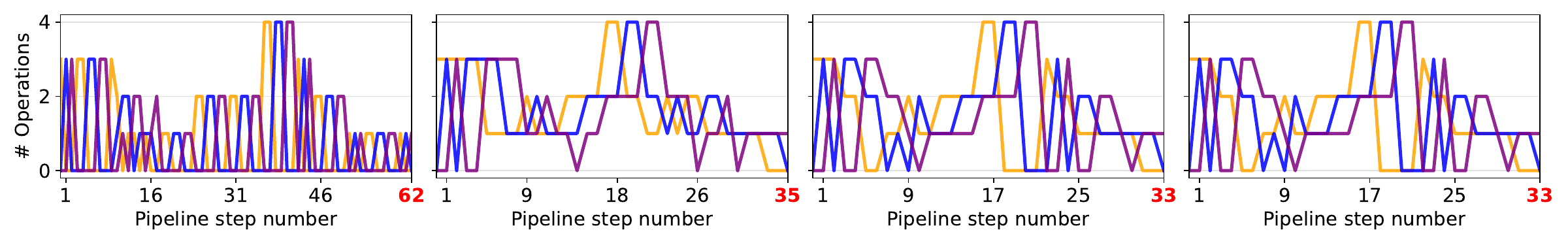}
    \caption{\texttt{adder\_n4}}
  \end{subfigure}
  \begin{subfigure}{\textwidth}
    \includegraphics[height=3cm,width=\columnwidth]{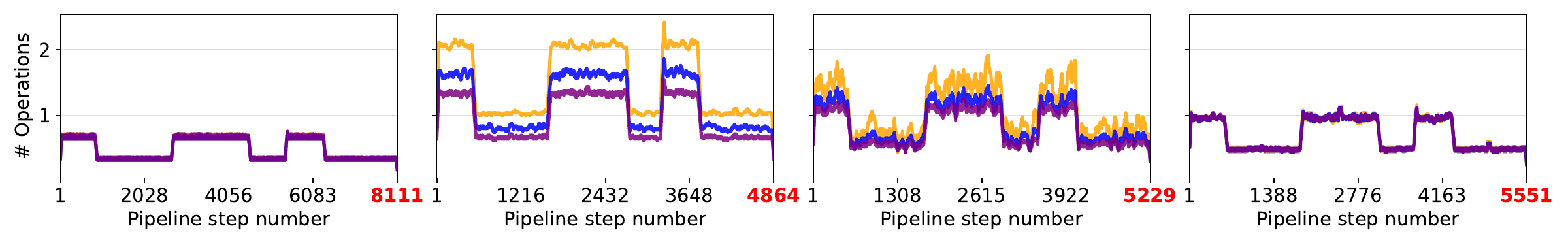}
    \caption{\texttt{wstate\_n3}}
  \end{subfigure}
  \caption{Pipeline stage utilization (smoothed) for representative benchmarks under four scheduling strategies. Without speculation, utilization is bursty. With speculation, all three subsystems sustain overlapping activity}
  \label{fig:occupancy_comparison}
\end{figure*}

Beyond reducing total cycles, speculation changes \emph{how} the three subsystems are utilized over time. Figure~\ref{fig:occupancy_comparison} plots the number of operations occupying each pipeline stage (Control, Execute, Decode) at every cycle for three representative benchmarks under all four strategies.

Under the no-speculation baseline, utilization is bursty: the three subsystems take turns being active rather than working in parallel. This is the temporal signature of the sequential scheduling problem identified in Section~\ref{sec:scheduling}. All three speculation strategies transform this profile: stage-occupancy traces become smoother, with all three subsystems active at most cycles simultaneously. For \texttt{adder\_n4}, total execution compresses from ${\sim}60$ to ${\sim}30$-$35$ steps; the same pattern holds for the larger benchmarks at proportionally greater scale. Speculation does not merely reduce total execution time; it increases the average number of in-flight operations per cycle, redistributing work across subsystems and converting idle time into useful computation.

\subsection{Circuit Structure Dependence}
The magnitude of speculation benefit depends on circuit structure. Circuits with high T gate density and deep dependency chains (e.g., \texttt{ising\_n26}, \texttt{knn\_n25}) show the largest absolute savings, because there are more opportunities for pipeline overlap across T gate boundaries. Circuits dominated by Clifford gates or with shallow DAGs show smaller but still positive savings, since Clifford operations don't lead to potential stalls.

Importantly, the relative ordering of strategies is stable across all circuit structures: \textit{Aggressive} $\geq$ \textit{Commute-Aware} $\geq$ \textit{T-Cautious} $>$ \textit{No Speculation}. Together, these results indicate that pipelining with speculation is a general-purpose optimization and holds favorable across circuit families, qubit counts, and T-gate densities, with the aggressive strategy as the best choice among the three we evaluated.

\section{Discussion}
\label{sec:discussion}
\emph{Relationship to prior work:} Litinski~\cite{Litinski_2019_lattice_surgery} observes that Clifford corrections can be tracked in software and commuted through subsequent operations, but does not develop this into a scheduling framework with explicit pipelining or speculation. Parallel window decoding works~\cite{Skoric_2023,viszlai2025swiper} optimize throughput within the \textit{Decode} stage; our framework operates one level above, pipelining across all three stages. Fowler and Gidney~\cite{fowler2019lowoverhead}, and Beverland et al.~\cite{qre2} treat operations as atomic scheduling units. Our approach is complementary since it can layer on top of any of these systems to reduce idle time across subsystems.

\emph{Pipeline stage length ratio:} Our evaluation assumes a 1:1:1 ratio for the three pipeline stages. In practice, the Decode stage is often the longest, particularly at large code distances where decoding algorithms scale superlinearly~\cite{ravi_2022}. A longer Decode stage widens the window during which speculation can overlap useful work with decoding, but reduces the relative benefit per speculated operation since Decode dominates total time. We expect speculation to remain beneficial under moderate asymmetry, with diminishing returns as the ratio grows. Quantifying this across realistic hardware parameters is an important direction for future work.

\emph{Assumptions:} We assume unbounded magic state availability, decoupling speculation strategy from factory scheduling. In practice, factory throughput constraints would introduce additional stalls, particularly for T-heavy circuits. Integrating our pipelined scheduler with factory-aware resource allocation~\cite{Sethi_2025} is a natural extension, and speculation could help partially mask factory latency by keeping subsystems busy during waits. Also, we model only independent injection failures for T gates. Correlated errors could amplify mis-prediction penalties for the aggressive strategy.

\section{Conclusion}
\label{sec:conclusion}
We have presented a pipelined execution framework for fault-tolerant quantum computation that decomposes each logical operation into Control, Execute, and Decode stages and overlaps them across the circuit dependency graph. We evaluated three speculation strategies: Aggressive, Commute-Aware, and T-Cautious, that allow successor operations to enter the pipeline before their predecessors have completed decoding.

All three strategies consistently reduce pipeline steps by 20-40\% compared to a no-speculation baseline. The Aggressive strategy achieves the largest savings on almost all benchmarks. Speculation also increases the average number of in-flight operations, smoothing utilization across the heterogeneous subsystems of a fault-tolerant quantum computer. More broadly, our results suggest that fault-tolerant quantum computers should be understood as heterogeneous multi-subsystem architectures. Pipelining and speculative execution--techniques that transformed classical processor design, have direct analogues in this setting. Our work is an initial exploration of this design space, intended to motivate further investigation into scheduling optimizations that remain largely unexplored in fault-tolerant quantum computation.

\section*{Acknowledgements}
The authors acknowledge the Texas Advanced Computing Center (\href{http://www.tacc.utexas.edu}{TACC}) at The University of Texas at Austin for providing computational resources that have contributed to the research results reported in this paper. This material is based upon work supported by the U.S. Department of Energy, Office of Science, Office of Advanced Scientific Computing Research, Accelerated Research in Quantum Computing under Award Number DE-SC0025633. This research was, in part, funded by the U.S. Government.  The views and conclusions contained in this document are those of the authors and should not be interpreted as representing the official policies, either expressed or implied, of the U.S. Government.

\bibliographystyle{IEEEtranS}
\bibliography{references}

\end{document}